\documentclass[twocolumn,10pt,journal]{IEEEtran}
\usepackage{epsfig}
\usepackage{amssymb}
\usepackage{varwidth}
\usepackage[cmex10]{amsmath}
\usepackage{algorithmic}
\usepackage{array}
\usepackage[usenames]{color}

\begin{document}
\title{ Robust Adaptive Beamforming Algorithms Based on the Constrained Constant Modulus Criterion}

\author{Lukas~Landau, Rodrigo~C.~de~Lamare, %~\IEEEmembership{Senior,~IEEE,}
and~Martin~Haardt %~\IEEEmembership{Senior,~IEEE}% <-this % stops a space
\thanks{R. C. de Lamare is with the Communications Research Group, Department
of Electronics, University of York, North Yorkshire, York Y010~5DD,
U.K. (e-mail: rcdl500@ohm.york.ac.uk).}% <-this % stops a space
\thanks{L. Landau and M. Haardt are with the Communications Research Laboratory, Ilmenau University of Technology, (Fachgebiet Nachrichtentechnik), D-98684 Ilmenau, Germany (e-mail: lukas.landau@tu-ilmenau.de, martin.haardt@tu-ilmenau.de)}% <-this % stops a space
}

\maketitle

\begin{abstract}
We present a robust adaptive beamforming algorithm based on the
worst-case criterion and the constrained constant modulus approach,
which exploits the constant modulus property of the desired signal.
Similarly to the existing worst-case beamformer with the minimum
variance design, the problem can be reformulated as a second-order
cone (SOC) program and solved with interior point methods. An
analysis of the optimization problem is carried out and conditions
are obtained for enforcing its convexity and for adjusting its
parameters. Furthermore, low-complexity robust adaptive beamforming
algorithms based on the modified conjugate gradient (MCG) and an
alternating optimization strategy are proposed. The proposed
low-complexity algorithms can compute the existing worst-case
constrained minimum variance (WC-CMV) and the proposed worst-case
constrained constant modulus (WC-CCM) designs with a quadratic cost
in the number of parameters. Simulations show that the proposed
WC-CCM algorithm performs better than existing robust beamforming
algorithms. Moreover, the numerical results also show that the
performances of the proposed low-complexity algorithms are
equivalent or better than that of existing robust algorithms,
whereas the complexity is more than an order of magnitude lower.
\end{abstract}

%\begin{keyword} Array signal processing \sep robust adaptive
%beamforming \sep array steering vector uncertainty \sep constrained
%constant modulus.
%\end{keyword}

\section{Introduction}

Beamforming has many applications in wireless communications, radar,
sonar, medical imaging, radio astronomy and other areas. One of the
most fundamental problems with adaptive beamforming {algorithms is
the} occurrence of mismatches between the presumed and actual signal
steering vector \cite{Trees02}. Practical circumstances like local
scattering, imperfectly calibrated arrays and imprecisely known
wavefield propagation conditions are the typical sources of these
mismatches and can lead to a performance degradation of the
conventional beamforming algorithms \cite{Li_book}. In the last
decades a number of robust approaches have been reported that
address this problem
\cite{Cox87,Gershman99,Vorobyov03,Stoica03,Li03,Li04,Lorenz05,Chen07,Hassanien08,Gershman10,Khabbazibasmenj10}.
These robust methods can be broadly categorized into two main
groups: techniques based on previous mismatch assumptions
\cite{Cox87,Gershman99,Vorobyov03,Li03,Li04,Lorenz05,Chen07} and
methods that estimate the mismatch or equivalently the actual
steering vector \cite{Stoica03,Li04,Hassanien08,Khabbazibasmenj10}.
A number of robust designs can be often cast as optimization
problems which end up in the so-called second-order cone (SOC)
program, which can be easily solved with interior point methods.
While those designs for beamformers are based on the minimum
variance criterion, it is possible to design them using a constant
modulus criterion \cite{{Lamare05}}, \cite{{Lamare08}}, which can
exploit prior knowledge about the desired signal and provide a
better performance.

The problem we are interested in solving in this paper is the design
of cost-effective adaptive robust beamforming algorithms. In
particular, we focus on the design of beamforming algorithms which
can exploit prior knowledge about the constant modulus property of a
desired signal and that can be implemented in an efficient way with
an appropriate modification of adaptive signal processing
algorithms. In the first part of this work the worst-case
optimization-based beamforming algorithm with the constant modulus
criterion (CCM) is developed. { In order to solve the robust
constrained constant modulus we apply an iterative reformulation of
the constant modulus cost function introduced in \cite{Chen04},
which is a local second-order approximation. Its derivation is based
on the assumption that previous computed weight vectors are close to
the solution, which is enforced by the additional constraint.}

We reformulate the problem as a SOC program in a similar fashion to
the approach adopted in \cite{Vorobyov03} and devise an adaptive
algorithm to adjust the parameters of the beamformer in time-varying
scenarios and that can exploit prior knowledge about the constant
modulus of the desired signal. An analysis of the optimization
problem is conducted and a condition which ensures convexity is
established. In addition, a study about the choice of the parameter
$\epsilon$ associated with the WC-CCM criterion is carried out. We
investigate the performance of the proposed WC-CCM algorithm via
simulations. The results show that the proposed WC-CCM algorithm
outperforms previously reported methods.

In the second part of this paper low-complexity robust adaptive
beamforming algorithms are developed.  While the
robust constraint is similar to that which is known from the
worst-case criterion, the algorithms are based on the modified
conjugate gradient (MCG) \cite{Wang10,Luenberger}and an alternating
optimization strategy that performs joint adjustment of the
constraint and the parameters of the beamformer. The joint
optimization strategy exploits previous computations and therefore
the computational complexity is reduced by more than an order of
magnitude from more than cubic $\mathcal O (M^{3.5})$ to quadratic
$\mathcal O (M^2)$ with the number of sensor elements
$M$ as compared to the worst-case
optimization-based approach. A low-complexity
approach is also developed for the minimum variance design which is
termed the robust constrained minimum variance modified conjugate
gradient (Robust-CMV-MCG) algorithm. The proposed low-complexity
algorithm for the constrained constant modulus design is termed
robust constrained constant modulus modified conjugate gradient
(Robust-CCM-MCG). While the Robust-CMV-MCG algorithm has a
performance equivalent to the worst-case optimization based
approach, the Robust-CCM-MCG algorithm which exploits the constant
modulus property of the desired signal performs better than existing
algorithms. We conduct a simulation study to investigate the
performance of the proposed low-complexity algorithms in a number of
situations of practical relevance.

This paper is organized as follows. The system model is described in
Section II. Section III reviews existing robust adaptive beamforming
algorithms. The proposed WC-CCM design is formulated in Section IV.
In Section V the SOC implementation and the adaptive algorithm are
described. An analysis of the optimization problem is given in
 {Section} VI, where a condition is found which
ensures convexity and relationships between the parameter $\epsilon$
and the signal-to-noise ratio (SNR) are established. In
 {Section} VII the corresponding low-complexity
solutions are presented. The simulation results are presented and
discussed in  {Section} VIII. Section IX gives the conclusion of
this work.

\section{System Model}
\label{sec:system} Let us consider a linear array of $M$ sensors
that receives signals from $D$ narrowband sources. The vector of
array observations $\boldsymbol{x}(i) \in \mathbb{C}^{M \times 1}$
 {at time instant $i$} can be modeled as
\begin{equation}
{\boldsymbol x}(i)=  {\boldsymbol A}( {\boldsymbol \theta} ) {\boldsymbol s} (i)+{\boldsymbol n}(i) ,
\end{equation}
where ${\boldsymbol \theta}=[\theta_1,..., \theta_{D} ]^T \in
\mathbb{R}^{D \times 1}  $ is the vector with the directions of
arrival (DoA) and $(.)^T$ stands for transpose, ${\boldsymbol A} (
\theta )=\left[ {\boldsymbol a}_1(\theta_1),... ,{\boldsymbol
a}_D(\theta_{D})    \right] \in \mathbb{C}^{M \times D}  $  is the
matrix containing the array steering vectors
${\boldsymbol a}_m(\theta_m) \in \mathbb{C}^{M \times 1}$ ,
for $m=1,...,D$. In the following $\theta_1$ is the
direction related to the desired user which is roughly known by the
system. The vector ${\boldsymbol s}(i) \in \mathbb{C}^{D \times 1}$
represents the uncorrelated sources. The vector ${\boldsymbol n}(i)
\in \mathbb{C}^{M \times 1} $ is the sensor noise, which is assumed
as zero-mean complex Gaussian. The true array steering vector is
assumed as ${\boldsymbol a}_1 (\theta_1) ={\boldsymbol a}(\theta_1)
+ {\boldsymbol e}$, where ${\boldsymbol e}$ is the mismatch vector
and ${\boldsymbol a}(\theta_1)$ is the presumed array steering
vector which is known by the system. In what follows, we will use
${\boldsymbol a} ={\boldsymbol a}(\theta_1)$. The output of the
beamformer is defined as
\begin{equation}
y(i)={\boldsymbol w}^H {\boldsymbol x}(i) \ , \label{output}
\end{equation}
where ${\boldsymbol w} \in \mathbb{C}^{M \times 1}$ is the complex
vector of beamforming weights. The notation $(.)^H$ stands for
Hermitian transpose. The signal-to-interference-plus-noise ratio
(SINR) is defined as
\begin{equation}
\mathrm{SINR}=\frac{{\boldsymbol w}^H \boldsymbol{R}_{s} {\boldsymbol w}}{{\boldsymbol w}^H  \boldsymbol{R}_{i+n}  {\boldsymbol w}} ,
\end{equation}
where $\boldsymbol{R}_{s}$ is the signal covariance matrix
corresponding to the desired user and $\boldsymbol{R}_{i+n}$ is
the interference-plus-noise covariance matrix.

\section{Robust Adaptive Beamforming: A Review}
\label{sec:review}

We review a few notable approaches to the design of robust adaptive
beamforming algorithms. The most common robust approach is the
so-called loaded sample matrix inversion (loaded-SMI) beamformer
\cite{Cox87}, which includes an additional diagonal loading to the
signal covariance matrix. The main problem is how to obtain the
optimal diagonal loading factor. Typically it is chosen as
$10~\sigma_n^2$, where $\sigma_n^2$ is the noise power
\cite{Cox87}. Another robust approach is given by
the eigen-based beamformer \cite{Feldman}. Here the presumed array
steering vector is replaced by its projection onto the
signal-plus-interferer subspace. The approach implies that the noise
subspace can be identified, which leads to a limitation in high SNR.
A similar method is given by the reduced-rank beamforming approach
\cite{Lamare10}, which avoids an eigen-decomposition and exploits
the low rank of the signal-plus-interferer subspace.  A different
robust beamforming strategy is considered by techniques based on
diagonal loading \cite{Vorobyov03,Li03,Lorenz05,Chen07}, which are
more advanced compared to \cite{Cox87}. In these techniques, the
algorithms determine a diagonal loading parameter which aims to
compensate for the mismatch by adding a suitable factor to the
diagonal of the covariance matrix of the input signal.

One of these methods is given by the popular worst-case performance
optimization-based beamformer \cite{Vorobyov03} which is based on
the constraint that the absolute value of the array response is
always greater  {than} or equal to a constant for all vectors that
belong to a predefined set of vectors in the neighborhood of the
presumed vector. In \cite{Vorobyov03} the set of vectors is a sphere
$\mathcal{A}= \left\{\boldsymbol{a}+\boldsymbol{e},
\left\|\boldsymbol{e}\right\|_{2}\leq \epsilon \right\}$, where the
norm of $\boldsymbol{e}$ is upper-bounded by $\epsilon$. The
corresponding optimization problem is given by
\begin{align}
& \min_{{\boldsymbol w}}  {\boldsymbol w}^H {\boldsymbol R}_{xx} {\boldsymbol w}  \notag \\
& {\rm s.~t.~} \left|{\boldsymbol w}^H \left({\boldsymbol a}+{\boldsymbol e}\right)\right| \geq \delta~~\rm{for~all} \left({\boldsymbol a}+{\boldsymbol e}\right)\in \mathcal{A}(\epsilon)
\end{align}
where ${\boldsymbol R}_{xx} = \mathrm{E}\left\{  \boldsymbol{x}
\boldsymbol{x} ^H   \right\}  $ is the covariance matrix of the
input signal. The problem can be transformed into
the following convex SOC problem:
\begin{equation}
\begin{split}
\min_{{\boldsymbol w}}  {\boldsymbol w}^H {\boldsymbol R}_{xx} {\boldsymbol w}  \ \ \
{\rm s.~t.~}  \operatorname{Re}  \left\{ {\boldsymbol w}^H {\boldsymbol a} \right\} - \delta \geq \epsilon \left\|{\boldsymbol w}\right\| _{2} \\
\operatorname{Im}  \left\{ {\boldsymbol w}^H {\boldsymbol a}
\right\} = 0  \,
\end{split}
\end{equation}
 {where the operator $\operatorname{Re}\{\cdot\}$
retains the real part of the argument and the operator$
\operatorname{Im}\{\cdot\}$ retains the imaginary part of the
argument}. It has been shown that this kind of beamforming technique
is related to the class of diagonal loading. In \cite{Lorenz05} the
set of vectors in the neighborhood can be ellipsoidal as well.

Another notable idea is the probability-constrained approach
\cite{Vorobyov07}. Here the constraint satisfies operational
conditions that are more likely to occur.
\begin{equation}
\begin{split}
\min_{{\boldsymbol w}}  {\boldsymbol w}^H {\boldsymbol R}_{xx} {\boldsymbol w}  \ \ \
{\rm s.~t.~} \mathrm{Pr}\left\{ \left|  {\boldsymbol w}^H \left({\boldsymbol a}+{\boldsymbol e}\right) \right| \geq \delta \right\}\geq p,
\end{split}
\end{equation}
where $\mathrm{Pr}$ denotes the probability operator and $p$ is the
desired probability threshold. Here different assumptions on the
statistical characteristics of the mismatch-vector $\boldsymbol{e}$
lead to different problem formulations. The solutions for the
Gaussian probability density function (pdf) case and the general
unknown pdf case have been developed in
\cite{Vorobyov07}.

Another class of robust methods includes those that estimate the
mismatch which have been reported in
\cite{Stoica03,Li04,Hassanien08,Khabbazibasmenj10}. The main idea
behind these approaches is to compute an estimate of the mismatch
and then subsequently use this information to obtain an estimate of
the actual steering vector. Recently developed approaches estimate
the mismatch vector based on sequential quadratic programming
\cite{Hassanien08} or based on semidefinite relaxation
\cite{Khabbazibasmenj10}.

All these beamformers are based on the minimum variance criterion.
We assume that a number of these beamformers can benefit from using
the CCM design criterion instead of the minimum variance one. Prior
work with the CCM design criterion includes the design of adaptive
beamformers \cite{Lei09,Wang10_avf} and receivers for spread
spectrum systems \cite{Lamare05,Lamare08,delamare2011}. The results
in the literature indicate that the CCM design has a superior
performance to those designs based on the minimum variance. In the
following, we develop a worst-case performance optimization-based
beamforming algorithm with the CCM design criterion. In addition, we
propose low-complexity robust beamforming algorithms.

\section{Proposed Worst-Case Optimization based Constant Modulus Design}
\label{sec:design}

The proposed robust beamformer is based on the worst-case approach.
In case of the minimum variance design it can be derived from the
following optimization problem
\begin{equation}
\begin{split}
\min_{{\boldsymbol w}}  {\boldsymbol w}^H {\boldsymbol R}_{xx} {\boldsymbol w}  \ \ \
{\rm s.~t.~}  \operatorname{Re}  \left\{ {\boldsymbol w}^H
{\boldsymbol a} \right\} - \delta \geq \epsilon \left\|{\boldsymbol w}\right\| _{2} \\
\operatorname{Im}  \left\{ {\boldsymbol w}^H {\boldsymbol a} \right\} = 0  \ ,
\end{split}
\end{equation}
where $\epsilon$ is the level of steering vector mismatch, which is
assumed as known a priori.  {The proposed beamformer uses the
constant modulus criterion, which exploits the constant modulus
property of the desired signal instead of the minimum variance
design criterion. To this end, we will assume that the signals
processed have a constant modulus property during the observation
time and the proposed algorithms are designed to exploit this
property.} The constant modulus cost function is defined by
\begin{align}
J=& \mathrm{E}\left\{ \left(\left|y(i) \right|^2- \gamma \right)^2
\right\} \notag \\
=& \mathrm{E}\left\{ \left( \boldsymbol{w}^H (i) \boldsymbol{x}(i)  \boldsymbol{x}^H(i)  \boldsymbol{w}(i)   - \gamma \right)^2  \right\} ,
\label{eq:CM_objective}
\end{align}
where $\gamma\geq0$ which is a parameter related  {to and should be
chosen according to the energy of the signal}. If the parameter
gamma is different then we need to choose the parameter delta of the
constraint to satisfy (30). . { By considering the approximation
strategy in \cite{Chen04}, that is, replacing in
(\ref{eq:CM_objective}) $\boldsymbol{w}^H (i) \boldsymbol{x}(i)$ by
$\boldsymbol{w}^H (i-1) \boldsymbol{x}(i)$, we obtain a modified
cost function which is a second-order local approximation
\begin{align}
\tilde{J}= \mathrm{E} \left\{ ( \boldsymbol{w}^H (i) \boldsymbol{x}(i) \boldsymbol{x}(i)^H   \boldsymbol{w}(i-1)    -\gamma )  ( \boldsymbol{w}^H (i-1) \boldsymbol{x}(i) \boldsymbol{x}(i)^H   \boldsymbol{w}(i)    -\gamma )    \right\}
\end{align}
This is a special case of the established general constant modulus
reformulation suggested in \cite{Chen04} whose validity has been
confirmed via computer experiments. Furthermore, it should also be
mentioned that the underlying assumption that the previous weight
vector is close to the solution is additionally enforced by the
direction constraint. Besides this strategy, there are similar
second-order approximation strategies in the literature that are
based on Taylor series expansion \cite{Meng09} approaches. By
discarding the constant term, the objective function is given by}
\begin{equation}
\hat{J}= {\boldsymbol w}^H \mathrm{E}\left\{  \left|y(i)\right|^2
{\boldsymbol x}(i) {\boldsymbol x}^H(i)\right\}{\boldsymbol w}  - 2
\gamma \operatorname{Re}   \left\{  {\boldsymbol w}^H
\mathrm{E}\left\{y^{*}(i) {\boldsymbol x}(i)  \right\} \right\},
\end{equation}
where $y(i)=\boldsymbol{w}^H(i-1) \boldsymbol{x}(i)$
 {denotes the output which assumes small variations of
the beamformer that allows the approximation $\boldsymbol{w}^H (i)
\boldsymbol{x}(i) \approx \boldsymbol{w}^H (i-1)
\boldsymbol{x}(i)$.} In combination with the worst-case constraint,
the proposed WC-CCM design can be cast as the following optimization
problem
\begin{equation}
\begin{split}
& \min_{{\boldsymbol w}}  {\boldsymbol w}^H {\boldsymbol R}_{a}
{\boldsymbol w}  - 2 \gamma~ \operatorname{Re} \left\{
{\boldsymbol d}^H {\boldsymbol w}   \right\}  \\
& {\rm s.~t.~}   {\boldsymbol w}^H {\boldsymbol a} - \delta \geq
\epsilon \left\|{\boldsymbol w}\right\| _{2} ~ {\rm and} ~
\operatorname{Im} \left\{ {\boldsymbol w}^H {\boldsymbol a}
\right\} = 0  \  , \label{prop_rob_opt}
\end{split}
\end{equation}
where ${\boldsymbol R}_{a} = \mathrm{E}\left\{  \left|y(i) \right|^2
{\boldsymbol x}(i) {\boldsymbol x}^H(i)\right\}$ and ${\boldsymbol
d}=\mathrm{E}\left\{ y^{*}(i) {\boldsymbol x}(i)  \right\}$, are
estimated from the previous snapshots which will be explained in the
next section.

\section{Proposed SOC Implementation and Adaptive Algorithm}
\label{sec:implementation}

In the first part of this section we show how to implement the SOC
program and in the second part we devise an adaptive algorithm to
adjust the weights of the beamformer according to the WC-CCM design.

\subsection{SOC Implementation}

In this subsection, inspired by the approach in
\cite{Vorobyov03}, we present a SOC implementation of the proposed
WC-CCM design. Introducing a scalar variable $\tau$, an equivalent
problem to (10) can be formulated
\begin{align}
\min_{\tau, {\boldsymbol w}}~~\tau~~
{\rm s.~t.~}
~~~& {\boldsymbol w}^H \boldsymbol{R}_{\textrm{ac}}^H \boldsymbol{R}_{\textrm{ac}}  {\boldsymbol w} - 2 \gamma~ \operatorname{Re} \left\{
{\boldsymbol d}^H {\boldsymbol w}   \right\}   \leq \tau  \notag   \\
& \mathrm{Re}\left\{{\boldsymbol w}^H {\boldsymbol a}\right\} - \delta \geq \epsilon \left\|{\boldsymbol w}\right\| _{2}  \notag \\
& \mathrm{Im}\left\{{\boldsymbol w}^H {\boldsymbol a}\right\}  = 0,
\end{align}
where ${\boldsymbol R}_{\textrm{ac}}^H {\boldsymbol
R}_{\textrm{ac}}={\boldsymbol R}_{a}$ is the Cholesky factorization.
Introducing the real-valued matrix and the real-valued vectors given
by  { $ \breve{{\boldsymbol R}}_{ac} =
\begin{bmatrix} \operatorname{Re} \left\{ {\boldsymbol R}_{ac}
\right\} & -\operatorname{Im} \left\{ {\boldsymbol R}_{ac} \right\}
\\ \operatorname{Im} \left\{ {\boldsymbol R}_{ac} \right\}
& \operatorname{Re} \left\{ {\boldsymbol R}_{ac} \right\}
\end{bmatrix}  \in \mathbb{R}^{(2M) \times (2M)}$, $\breve{{\boldsymbol d}}
 =  [ \operatorname{Re}\left\{ {\boldsymbol d}  \right\}^T ,
\operatorname{Im} \left\{ {\boldsymbol d}  \right\}^T  ]^T \in
\mathbb{R}^{(2M) \times 1}$, $\breve{{\boldsymbol a}}  =  [
\operatorname{Re} \left\{{\boldsymbol a}\right\}^T ,
\operatorname{Im} \left\{{\boldsymbol a}\right\}^T ]^T \in
\mathbb{R}^{(2M) \times 1}$, $\bar{{\boldsymbol a}}  = [
\operatorname{Im} \left\{{\boldsymbol a}\right\}^T ,
-\operatorname{Re} \left\{{\boldsymbol a}\right\}^T ]^T \in
\mathbb{R}^{(2M) \times 1}$, $\breve{{\boldsymbol w}}  = [
\operatorname{Re}\left\{ {\boldsymbol w}  \right\}^T ,
\operatorname{Im} \left\{ {\boldsymbol w}  \right\}^T  ]^T \in
\mathbb{R}^{(2M) \times 1}$.} The problem can be rewritten as
\begin{align}
\min_{\tau, \breve{{\boldsymbol w}}}~~\tau~~
{\rm s.~t.~}
~~~& \breve{{\boldsymbol w}}^T \breve{{\boldsymbol R}}_{ac} ^T \breve{{\boldsymbol R}}_{ac}   \breve{{\boldsymbol w}} -  2 \gamma ~ \breve{{\boldsymbol d}}^T \breve{{\boldsymbol w}}  \leq \tau  \notag   \\
& \breve{{\boldsymbol w}}^T \breve{{\boldsymbol a}} - \delta \geq \epsilon \left\|\breve{{\boldsymbol w}}\right\| _{2}  \notag \\
& \breve{{\boldsymbol w}}^T \bar{{\boldsymbol a}}  = 0.
\end{align}
 {The quadratic constraint can be converted into an
equivalent SOC constraint because the convexity of the optimization
problem can be enforced as will be shown in the next section.} This
leads to the following optimization problem
\begin{align}
\min_{\tau, \breve{{\boldsymbol w}}}~~\tau~~
{\rm s.~t.~}
~~~&  \frac{1}{2} +  \gamma \breve{{\boldsymbol d}}^T \breve{{\boldsymbol w}} + \frac{\tau}{2}   \geq     \left\|  \begin{bmatrix}   \frac{1}{2} - \gamma \breve{{\boldsymbol d}}^T \breve{{\boldsymbol w}}  - \frac{\tau}{2}   \\    \breve{{\boldsymbol R}}_{\textrm{ac}} \breve{{\boldsymbol w}}    \end{bmatrix}  \right\|_2      \notag   \\
& \breve{{\boldsymbol w}}^T \breve{{\boldsymbol a}} - \delta \geq \epsilon \left\|\breve{{\boldsymbol w}}\right\| _{2}  \notag \\
& \breve{{\boldsymbol w}}^T \bar{{\boldsymbol a}}  = 0.
\end{align}
Let us define
\begin{eqnarray*}
{\boldsymbol p} & = & [ 1, {\boldsymbol 0}^T ]^T   \in \mathbb{R}^{(2M+1) \times 1}  \\
{\boldsymbol u} & = & [ \tau, \breve{{\boldsymbol w}}^T  ]^T  \in \mathbb{R}^{(2M+1) \times 1}  \\
{\boldsymbol f} & = & [ 1/2 , 1/2 ,  {\boldsymbol 0}^T,-\delta,{\boldsymbol 0}^T,0 ]^T  \in \mathbb{R}^{(4M+4) \times 1}        \\
{\boldsymbol F}^T & = & \begin{bmatrix}   \frac{1}{2} & \gamma \breve{{\boldsymbol d}}^T \\ -\frac{1}{2} & -  \gamma \breve{{\boldsymbol d}}^T \\  {\boldsymbol 0} & \breve{{\boldsymbol R}}_{\textrm{ac}} \\ 0 & \breve{{\boldsymbol a}} \\  {\boldsymbol 0} & \epsilon {\boldsymbol I} \\ 0 & \bar{{\boldsymbol a}}  \end{bmatrix}  \in \mathbb{R}^{(4M+4) \times (2M+1)},
\end{eqnarray*}
where $\boldsymbol{I}$ is a $2M \times 2M$ identity matrix and
${\boldsymbol 0}$ is a vector of zeros of compatible dimensions.
Finally, the problem can be formulated as the dual form of the SOC
problem (equivalent to (8) in \cite{SeDuMi})
\begin{align}
\min_{{\boldsymbol u}}~~{\boldsymbol p}^T {\boldsymbol u} ~~
 & {\rm s.~t.~}  \notag  \\
~~~ & {\boldsymbol f}+ {\boldsymbol F}^T {\boldsymbol u}~~ \in  \mathrm{SOC}_{1}^{2M+2} \times  \mathrm{SOC}_{2}^{2M+1}   \times \{0 \},
\end{align}
where ${\boldsymbol f}+ {\boldsymbol F}^T {\boldsymbol u}$ describes
a SOC with a dimension $2M+2$, a SOC with a dimension $2M+1$
 and $\{0\}$ is the so-called zero cone that
determines the hyperplane due to the equality constraint
$\breve{{\boldsymbol w}}^T \bar{{\boldsymbol a}}  = 0$. Finally, the
weight vector of the beamformer $\boldsymbol{w}$ can be retrieved in
the  {form ${\boldsymbol w}= \left[ {\boldsymbol
u}_2,...,{\boldsymbol u}_{M+1}\right]^T  + j \left[ {\boldsymbol
u}_{M+2},...,{\boldsymbol u}_{2M+1}    \right]^T$}. Alternatively
(10) can be solved by using \cite{cvx}, which transforms it
automatically into an appropriate form.

\subsection{Adaptive Algorithm}

It has already been mentioned that the optimization problem
corresponding to the WC-CCM algorithm design is solved iteratively.
As a result, the underlying optimization problem is to be solved
periodically. In this case the proposed adaptive algorithm solves it
at each time instant. For the adaptive implementation we use an
exponentially decayed data window for the estimation of
${\boldsymbol R}_{a}$ and ${\boldsymbol d}$ given by
\begin{align}
\hat{{\boldsymbol R}}_{a}(i)=\mu \hat{{\boldsymbol R}}_{a}(i-1)+ \left|y(i)\right|^2  {\boldsymbol x}(i){\boldsymbol x}^H(i) \\
\hat{\boldsymbol d}(i)=\mu \hat{\boldsymbol d}(i-1) + {\boldsymbol x}(i) y^{*}(i),
\end{align}
where $0< \mu<1$ is the forgetting factor. Each
iteration includes a Cholesky factorization and also a
transformation into a real valued problem. Finally, the problem is
formulated in the dual form of the SOC problem and solved with
SeDuMi \cite{SeDuMi}. The structure of the adaptive
algorithm is summarized in Table \ref{tab:wcccm}. Compared to the
algorithm based on the minimum variance constraint, the proposed
algorithm increases the dimension of the first SOC from $2M+1$ to
$2M+2$.

\section{Analysis of the Optimization Problem}

In this section we analyze the optimization problem associated with
the design of the proposed robust WC-CCM beamformer. For the purpose
of analysis, we rely on the equality of the robust constraint described
in (\ref{prop_rob_opt}). In particular,
we derive a sufficient condition for enforcing the convexity of the
proposed WC-CCM beamformer design as a function of the power of the
desired signal. We also provide design guidelines for the adjustment
of the parameter $\epsilon$ in the optimization problem.

\subsection{Convexity of the Optimization Problem}

The objective function for the constant modulus design criterion is
\begin{align}
J_{\textrm{cm}}= \mathrm{E}\left\{ \left(\left|y(i)\right|^2- \gamma
\right)^2   \right\}.
\end{align}
To ensure that the constraint $\boldsymbol{w}^H
\boldsymbol{a}=\delta+\epsilon~\left\| \boldsymbol{w} \right\|_{2}$
is fulfilled, the beamformer $\boldsymbol{w}$ is
replaced by
\begin{align}
\boldsymbol{w}=   \frac{ \boldsymbol{a} } {M}~\left(\delta+ \epsilon
\left\| \boldsymbol{w}   \right\|_{2}   \right)+ \boldsymbol{B}
\boldsymbol{z},\label{rel_w}
\end{align}
where the columns of $\boldsymbol{B}$ are unitary and span the null
space of $\boldsymbol{a}^H$, $\boldsymbol{z} \in
\mathbb{C}^{M-1 \times 1}$ and $\boldsymbol{a}^H \boldsymbol{a} = M$. To obtain a function which does not
depend on $\left\| \boldsymbol{w}\right\|_{2}$, we compute the
squared norm of (\ref{rel_w}) and obtain the following quadratic
equation to be solved:
\begin{equation}
\left\| \boldsymbol{w}\right\|_{2}=\tau=\sqrt{ \frac{1}{M} \left(\delta+\epsilon~\tau \right)^2 + \boldsymbol{z}^H \boldsymbol{z} }  \label{eqwn}
\end{equation}
Since the norm is greater than zero the following holds
\begin{align}
\tau=\frac{\epsilon~\delta}{M-\epsilon^2}+\sqrt{ \frac{M \boldsymbol{z}^H \boldsymbol{z}+\delta^2  }{M-\epsilon^2}  +\left(\frac{\epsilon~\delta}{M-\epsilon^2}\right)^2 }. \label{eqtau}
\end{align}
Therefore, by inserting (\ref{eqwn}) and (\ref{eqtau}) in (\ref{rel_w}) , the resulting weight vector $\boldsymbol{w}$ is a function of
$\boldsymbol{z}$ as described by
\begin{align}
{\footnotesize{
\boldsymbol{w}\left( \boldsymbol{z}  \right)=  \frac{ \boldsymbol{a} } {M}~\left(\delta+    \frac{\epsilon^2~\delta}{M-\epsilon^2}+\epsilon~\sqrt{ \frac{M \boldsymbol{z}^H \boldsymbol{z}+\delta^2  }{M-\epsilon^2}  +\left(\frac{\epsilon~\delta}{M-\epsilon^2}\right)^2 }    \right)+ \boldsymbol{B} \boldsymbol{z} }}
\end{align}
Replacing the $\boldsymbol{w}$ in the objective function leads to an
equivalent problem to the original:
\begin{align}
J = \mathrm{E}\left\{  \left| y(i) \right|^2-\gamma  \right\} = \mathrm{E}
\left\{ \left[ \boldsymbol{w}^H\left( \boldsymbol{z}  \right)
\boldsymbol{x} \boldsymbol{x}^H \boldsymbol{w} \left( \boldsymbol{z}
\right)                    - \gamma \right]^2  \right\}
\end{align}
The function above is convex if the Hessian $\boldsymbol{H}=
\frac{\partial }{\partial \boldsymbol{z}^H}\left( \frac{\partial J}
{\partial \boldsymbol{z}}  \right)  $ is positive semi-definite. The
Hessian corresponding to the objective function is given by
\begin{align}
\boldsymbol{H} =  2 \frac{\partial }{\partial \boldsymbol{z}^H}\left( \mathrm{E} \left\{  \left| y \right|^2-\gamma  \right\} \right)   \frac{\partial }{\partial \boldsymbol{z}}\left( \mathrm{E} \left\{  \left| y \right|^2-\gamma  \right\} \right) \notag \\ + 2~\mathrm{E} \left\{  \left| y \right|^2-\gamma  \right\}  \frac{\partial} {\partial \boldsymbol{z}^H}  \frac{\partial} {\partial \boldsymbol{z}} \left( \mathrm{E} \left\{ \left| y \right|^2-\gamma \right\}  \right)
\label{eq:hessian_1}
\end{align}
Since it is the product of a vector multiplied with its Hermitian
transposed the first term in (\ref{eq:hessian_1}), is positive
semi-definite. While it is assumed that $\mathrm{E} \left\{  \left|
y(i) \right|^2-\gamma  \right\} \geq 0 $ the positive
semi-definiteness of $\boldsymbol{H}_{2} =\frac{\partial} {\partial
\boldsymbol{z}^H} \frac{\partial} {\partial \boldsymbol{z}} \left(
\mathrm{E} \left\{ \left| y \right|^2-\gamma \right\}  \right)$
still needs to be shown. It can be expressed as a sum
 $\boldsymbol{H}_{2}= \sum_{k=1}^{4}
\boldsymbol{H}_{2,k}$ and is given by
\begin{align}
& \boldsymbol{H}_{2} = \notag \\
&  \mathrm{E}\Bigg\{  \left( - \frac{\epsilon}{2}  \left( \frac{1}{\sqrt{\alpha}}    \right)  \right)^3 \left( \frac{M}{M-\epsilon^2} \right)^2 \operatorname{Re}  \left\{  \xi  \right\}  \boldsymbol{z} \boldsymbol{z}^H  \notag \\
&+ \epsilon \frac{1}{\sqrt{\alpha}} \left( \frac{M}{M-\epsilon^2} \right) \operatorname{Re}  \left\{  \xi  \right\}  \boldsymbol{I}_{M-1}  \notag \\
&+ \frac{\epsilon}{2} \frac{1}{\sqrt{\alpha}} \left( \frac{M}{M-\epsilon^2} \right) \frac{\boldsymbol{a}^H}{M}  \boldsymbol{x} \boldsymbol{x}^H \frac{\boldsymbol{a}}{M} \frac{\epsilon}{2} \frac{1}{\sqrt{\alpha}} \left( \frac{M}{M-\epsilon^2} \right) \boldsymbol{z} \boldsymbol{z}^H  \notag \\
&+\left( \frac{\epsilon}{2} \frac{1}{\sqrt{\alpha}} \left( \frac{M}{M-\epsilon^2} \right) \boldsymbol{z}\frac{\boldsymbol{a}^H}{M}+\boldsymbol{B}^H\right)  \boldsymbol{x} \boldsymbol{x}^H \left(\frac{\boldsymbol{a}}{M} \frac{\epsilon}{2} \frac{1}{\sqrt{\alpha}} \left( \frac{M}{M-\epsilon^2} \right) \boldsymbol{z}^H +\boldsymbol{B} \right)
 \Bigg\},
\end{align}
where $\alpha = \left(\frac{M \boldsymbol{z}^H \boldsymbol{z}+\delta^2  }{M-\epsilon^2}  +\left(\frac{\epsilon~\delta}{M-\epsilon^2}\right)^2 \right)$, $\beta=\left(\delta+    \frac{\epsilon^2~\delta}{M-\epsilon^2}+\epsilon~\sqrt{\alpha }\right)$ and $ \xi=\frac{\boldsymbol{a}^H}{M}  \boldsymbol{x} \boldsymbol{x}^H \left( \beta  \frac{ \boldsymbol{a}}{M}+\boldsymbol{B}\boldsymbol{z}   \right) $.
To show that $\boldsymbol{H}_2$ is positive semidefinite the following steps are made. Here it is assumed that
\begin{align}
\operatorname{Re}  \left\{  \xi  \right\}   = \operatorname{Re}  \left\{  \frac{\boldsymbol{a}^H}{M}  \boldsymbol{x} \boldsymbol{x}^H \left( \beta  \frac{ \boldsymbol{a}}{M}+\boldsymbol{B}\boldsymbol{z}   \right)   \right\}  \geq0 .
\end{align}
This assumption is reasonable as far as the term $\boldsymbol{x}^H
\boldsymbol{B}\boldsymbol{z}$ is basically the compensating term of
the unwanted contribution of $\boldsymbol{x}^H \left( \beta  \frac{
\boldsymbol{a}}{M}   \right)$. Under this condition all terms in the
sum of $\boldsymbol{H}_{2}$ are positive
semi-definite except the first term
$\boldsymbol{H}_{2,1}$. The inequality $\boldsymbol{v}^H
(\boldsymbol{H}_{2,2})  \boldsymbol{v}\geq \boldsymbol{v}^H
(-\boldsymbol{H}_{2,1})  \boldsymbol{v}  \ \  \forall \boldsymbol{v}$  is a sufficient condition
to ensure positive semi-definiteness which is described as
\begin{align}
\boldsymbol{v}^H        \epsilon \frac{1}{\sqrt{\alpha}} \left( \frac{M}{M-\epsilon^2} \right) \operatorname{Re}  \left\{  \xi \right\} \boldsymbol{I}_{M-1}                  \boldsymbol{v}  \notag \\
\geq
\boldsymbol{z}^H   \left(  \frac{\epsilon}{2}  \left( \frac{1}{\sqrt{\alpha}}    \right)  \right)^3 \left( \frac{M}{M-\epsilon^2} \right)^2 \operatorname{Re}  \left\{  \xi \right\}  \boldsymbol{z} \boldsymbol{z}^H  \boldsymbol{z} \notag \\
\geq  \boldsymbol{v}^H   \left(  \frac{\epsilon}{2}  \left( \frac{1}{\sqrt{\alpha}}    \right)  \right)^3 \left( \frac{M}{M-\epsilon^2} \right)^2 \operatorname{Re}  \left\{  \xi \right\} \boldsymbol{z} \boldsymbol{z}^H  \boldsymbol{v} ,
\end{align}
where $\boldsymbol{v}$ is any vector with the same norm of
$\boldsymbol{z}$ and $\boldsymbol{z}^H (-\boldsymbol{H}_{2,1})  \boldsymbol{z}$ is intruduced as the upper bound for $\boldsymbol{v}^H
(-\boldsymbol{H}_{2,1})  \boldsymbol{v}$. Since
$\boldsymbol{z}^H \boldsymbol{z}=\boldsymbol{v}^H \boldsymbol{v}$,
the inequality can be reduced to
\begin{align}
2 \alpha \geq \left( \frac{M \boldsymbol{z}^H \boldsymbol{z}}{M-\epsilon^2}  \right).
\end{align}
Replacing $\alpha$ gives the proof for positive semi-definiteness
\begin{equation}
2 \left( \frac{M \boldsymbol{z}^H
\boldsymbol{z}+\delta}{M-\epsilon^2}+\left(\frac{\epsilon~\delta}{M-\epsilon^2}\right)^2
\right) \geq \left( \frac{M \boldsymbol{z}^H
\boldsymbol{z}}{M-\epsilon^2}  \right),
\end{equation}
which is always true. To ensure that
$\mathrm{E}\left\{\left|y\right|^2-\gamma\right\} \geq 0$ it can be
assumed that $\boldsymbol{w}^H
\left(\boldsymbol{a}+\boldsymbol{e}\right)\geq \delta $, where
$\boldsymbol{e}$ is the array steering vector mismatch. Therefore,
\begin{align}
\gamma \leq   \delta~\mathrm{E} \left\{ \left|s_{1}\right|^2\right\}
\label{conv_cond}
\end{align}
is a sufficient condition to enforce convexity, where
$\left|s_{1}\right|^2$ is the power of the desired user.
 {Therefore, the parameter gamma should be chosen such
that the convexity condition given in (\ref{conv_cond}) is
satisfied.}

\subsection{Adjustment of the Design Parameter $\epsilon$}

Let us define the beamforming weight vector as
\begin{equation}
\label{eqn:epsilon_eq0}
{\boldsymbol w} = c~{\boldsymbol a}/M + {\boldsymbol b},
\end{equation}
where $c$ is a scalar, and ${\boldsymbol b}$ is orthogonal to ${\boldsymbol a}$.
%The structure of ${\boldsymbol b}$ depends on the interference circumstances.
Using it with the worst-case constraint leads to
\begin{equation}
c-\delta \geq  \epsilon~\sqrt{   \frac{c^2}{M}+ {\boldsymbol b}^H {\boldsymbol b} }.
\end{equation}
From the above inequality the following relation
holds
\begin{equation}
\label{eqn:epsilon_eq} c-\delta \geq  \epsilon~\sqrt{ \frac{c^2}{M}+
{\boldsymbol b}^H {\boldsymbol b} } \geq \epsilon~\sqrt{
\frac{c^2}{M} }.
\end{equation}
Rewriting the relation shows that $c$ tends to
infinity when $\epsilon$ is close to $\sqrt{M}$
\begin{equation}
c \geq \frac{\delta}{1-\epsilon / \sqrt{M}}.
\end{equation}
In addition, it is mentioned in \cite{Li03} that for $\left\|
\boldsymbol{a} \right\|_{2} \leq \epsilon $
%and $\left\| \boldsymbol{a} \right\|_{2} = \sqrt{M}$
there is no $\boldsymbol{w}$ that satisfies the constraint.
By rewriting the inequality in
(\ref{eqn:epsilon_eq}), we obtain
\begin{equation}
c \geq \frac{M~\delta}{M-\epsilon^2}+\sqrt{ \frac{  M \epsilon^2 \boldsymbol{b}^H \boldsymbol{b}-M~\delta }{M-\epsilon^2}+ \left(\frac{M~\delta}{M-\epsilon^2}\right)^2 }
\end{equation}

Now by assuming that $\epsilon \approx \sqrt{M}$ and
strictly less than $\sqrt{M}$, then we have $c \gg \delta$. In that case,
the inequality in (\ref{eqn:epsilon_eq}) can be
rewritten as
\begin{equation}
c \geq \epsilon \sqrt{ \frac{ c^2 }{M}+ {\boldsymbol b}^H {\boldsymbol b} },
\end{equation}
or equivalently as
\begin{equation}
\label{eqn:epsilon_eq2}
\frac{\boldsymbol{b}^H \boldsymbol{b}}{c^2} \leq \frac{1}{\epsilon^2}-\frac{1}{M} .
\end{equation}
As a result of (\ref{eqn:epsilon_eq2}), the choice
of $\epsilon$ affects the ratio between the components of the weight
vector defined by (\ref{eqn:epsilon_eq0}), which can become
negligible. This corresponds to $\boldsymbol{w} \approx
c~\boldsymbol{a}/M$ and an equivalent diagonal loading which is
above the level of the interference. Hence, the diagonal loading can
be chosen by an appropriate procedure if $\epsilon$ is chosen in the
allowed interval $[0, \sqrt{M}]$, where the constraint can be
enforced. Obviously, in the case of $\epsilon$ being close to
$\sqrt{M}$ the ratio $\frac{\boldsymbol{b}^H \boldsymbol{b}}{c^2}$
tends to a small value, which can lead to a performance degradation.
The ratio is small for low SNR values, and this is caused by the
assumption that the additional noise appears as a diagonal loading
in the signal covariance matrix and this eventually decreases
$ \left\| \boldsymbol{b} \right\|$. This means that the relation in
(\ref{eqn:epsilon_eq2}) and its penalty has a more significant
impact on the performance for higher SNR values. As a consequence
our suggestion is to choose $\epsilon$ with respect to the SNR as
well as with respect to the assumed mismatch level. This will be investigated in the simulations (see Fig.\ref{fig:fig2})

\section{Low-Complexity Algorithms using the Modified Conjugate Gradient}
\label{sec:joint-mcg}

The existing algorithms which use the worst-case optimization-based
constraint do not take advantage of previous computations as the
conventional SMI beamforming algorithm solved by the modified
conjugate method (MCG) algorithm or the recursive-least-squares
(RLS) algorithm in the so-called \textit{on-line} mode. For this
reason, the existing robust beamforming algorithms are not suitable
for low-complexity implementations and are unable to track
time-varying signals.

In this section a robust constraint is shown which is just slightly
different compared to the worst-case optimization-based approaches.
As a result the corresponding optimization problem is a
quadratically constrained quadratic program (QCQP) instead of a
second order cone (SOC) program. It is shown how to solve the
problem with a joint optimization strategy. The method includes a
system of equations which is solved efficiently with a modified
conjugate gradient algorithm and an alternating optimization
strategy \cite{Niesen}. As a result, the computational complexity is
reduced from more than cubic $\mathcal O (M^{3.5})$ to quadratic
$\mathcal O (M^2)$ with the number of sensor elements, while the
SINR performance is equivalent to the worst-case optimization-based
approach. The proposed method is presented in the robust constrained
minimum variance design using the modified conjugate gradient method
(RCMV-MCG) and in the robust constrained constant modulus design
using the modified conjugate gradient method (RCCM-MCG), which
exploits the constant modulus property of the desired signal.

\subsection{Proposed Design and Joint Optimization Approach}

\begin{figure}[!htb]
\begin{center}
\def\epsfsize#1#2{0.65\columnwidth}
\epsfbox{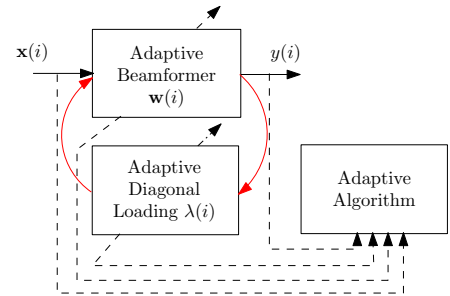} \caption{Proposed adaptive scheme with
alternating updates of beamforming weights and diagonal loading.}
\label{fig:fig0}
\end{center}
\end{figure}

In this part, we detail the main steps of the proposed design and
the low-complexity algorithms as well as the joint optimization
approach that is employed to compute the parameters of the adaptive
robust beamformer and the diagonal loading. Specifically, the
proposed algorithms are based on an alternating optimization
strategy \cite{Niesen} that updates the beamformer ${\boldsymbol
w}(i)$ while the diagonal loading $\lambda(i)$ is fixed and then
updates $\lambda(i)$ while ${\boldsymbol w}(i)$ is held fixed. The
algorithm is illustrated in Fig. \ref{fig:fig0}.

%\begin{figure}[!htb]
%\begin{center}
%\def\epsfsize#1#2{0.9\columnwidth}
%\epsfbox{adap_scheme.eps} \caption{Proposed adaptive scheme with
%alternating updates of beamforming weights and diagonal loading.}
%\label{fig:fig0}
%\end{center}
%\end{figure}

Since the joint optimization of the parameters ${\boldsymbol w}(i)$
and $\lambda(i)$ is not a convex optimization problem, the first
question that arises is whether the proposed algorithms will
converge to their global minima. The proposed algorithms have been
widely tested and we have not observed problems with
local minima. This is corroborated by the recent results reported
in \cite{Niesen} that shows that alternating optimization techniques
similar to that proposed here converge to the global optimum
provided that typical assumptions used for adaptive algorithms such
as step size values, forgetting factors and the
statistical independence of the noise and the source data processed
hold.

\subsubsection{Robust Constrained Minimum Variance Design}

The proposed low-complexity beamforming algorithms are related to
the worst-case approach \cite{Vorobyov03}. In order to obtain a
design which can be solved with a low complexity,
the robust constraint reported in \cite{Vorobyov03} is modified.
According to \cite{Lorenz05} it is sufficient to
 {use} the real part of the constraint. In addition,
it is assumed that the use of $\tilde{\epsilon}  \left\|{\boldsymbol
w}\right\|^{2}_{2}$ instead of $\epsilon \left\|{\boldsymbol
w}\right\|_{2}$ from the conventional constraint has a comparable
impact. Finally, the proposed design criterion for the minimum
variance case is
\begin{equation}
\begin{split}
\min_{{\boldsymbol w}}  {\boldsymbol w}^H {\boldsymbol R}_{xx} {\boldsymbol w}, \ \ \
{\rm s.~t.~} \operatorname{Re} \left\{ {\boldsymbol w}^H {\boldsymbol a} \right\} - \delta \geq \tilde{\epsilon}  \left\|{\boldsymbol w}\right\|^{2}_{2} .
\end{split}
\end{equation}
Using the method of Lagrange multipliers gives
\begin{equation}
{\mathcal L}_{\textrm{CMV}} \left( {\boldsymbol w},\lambda    \right) =  {\boldsymbol w}^H {\boldsymbol R}_{xx} {\boldsymbol w}  + \lambda \left[ \tilde{\epsilon} \ {\boldsymbol w}^H{\boldsymbol w} -  \operatorname{Re} \left\{ {\boldsymbol w}^H {\boldsymbol a} \right\} + \delta \right],
\label{eq:lagrange_rcmv}
\end{equation}
where $\lambda$ is the Lagrange multiplier.
Computing the gradient of (\ref{eq:lagrange_rcmv}) with respect to ${\boldsymbol w}^{*}$, and equating it to zero leads to
\begin{equation}
{\boldsymbol w} =   \left(  {\boldsymbol R}_{xx}  +  \tilde{\epsilon} \lambda  {\boldsymbol I}    \right)^{-1} \lambda{\boldsymbol a}/2 .
\label{eq:w_rcmv}
\end{equation}
Since there is no known method in the literature that can obtain the
Lagrange multiplier in a closed form, here it is proposed a strategy
to adjust both the beamformer ${\boldsymbol w}$ and
the Lagrange multiplier in an alternating fashion. In this joint
optimization the Lagrange multiplier is interpreted as a penalty
factor and the condition $\lambda>0$ holds all the time. The
adjustment increases the penalty factor when the constraint is not
fulfilled and decreases it otherwise. To this end,
we devise the following algorithm
\begin{equation}
\lambda (i)  = \lambda(i-1) + \mu_{\lambda}  \left(     \tilde{\epsilon}  \left\|{\boldsymbol w}(i)\right\|^{2}_{2} -  \operatorname{Re} \left\{ {\boldsymbol w}(i)^H {\boldsymbol a} \right\} + \delta      \right) \ ,
\end{equation}
where $\mu_{\lambda}$ is the step size. In addition, it is
reasonable to define boundaries for the update term.

{  In order to obtain an operation range for the parameter
$\tilde{\epsilon}$ the weight vector can be expressed as
$\boldsymbol{w}=\frac{ \boldsymbol{a}}{M}+ \boldsymbol{b}$.
Rearranging the constraint function leads to the inequality
\begin{align}
\tilde{\epsilon}\leq \frac{c-\delta}{ \frac{1}{M} c^2   +\boldsymbol{b}^H \boldsymbol{b} }  \leq  M \frac{c-\delta}{  c^2  } \leq \frac{M}{2},
\end{align}
which clearly indicates that there is no solution for
$\tilde{\epsilon}>\frac{M}{2}$. From our experiments we know that
the parameter has to be chosen significantly smaller. }

\subsubsection{Robust Constrained Constant Modulus Design}

In case of constant modulus signals it has been shown that the
constant modulus design performs better than the minimum variance
design \cite{Lamare05}, \cite{Lamare08}. Similarly, the robust
approach can be combined with the constrained constant modulus
criterion. The corresponding optimization problem for the
iteratively solved constant modulus objective function can be cast
as
\begin{align}
\min_{{\boldsymbol w}}     {\boldsymbol w}^H {\boldsymbol R}_{\textrm{a}} {\boldsymbol w}-2\gamma \operatorname{Re}\left\{ {\boldsymbol w}^H  {\boldsymbol d}   \right\}, \\
{\rm s.~t.~} \operatorname{Re} \left\{ {\boldsymbol w}^H {\boldsymbol a} \right\} - \delta \geq \tilde{\epsilon} \left\|{\boldsymbol w}\right\|^{2}_{2}
\end{align}
Using the method of Lagrange multipliers gives
\begin{align}
{\mathcal L}_{\textrm{CCM}} \left( {\boldsymbol w},\lambda    \right)  = & {\boldsymbol w}^H {\boldsymbol R}_{\textrm{a}} {\boldsymbol w} -2~\gamma \operatorname{Re}\left\{ {\boldsymbol w}^H  {\boldsymbol d}   \right\} \notag \\
&  + \lambda \left[ \tilde{\epsilon} \ {\boldsymbol w}^H{\boldsymbol w} -  \operatorname{Re} \left\{ {\boldsymbol w}^H {\boldsymbol a} \right\} + \delta \right].
\label{eq:lagrange_rccm}
\end{align}
Computing the gradient of (\ref{eq:lagrange_rccm}) with respect to
${\boldsymbol w}^{*}$, and equating it to zero leads to
\begin{align}
{\boldsymbol w}=\left[ {\boldsymbol R}_{\textrm{a}}+
\tilde{\epsilon} \lambda  {\boldsymbol I}  \right]^{-1} \left[
\gamma {\boldsymbol d} +\lambda {\boldsymbol a} / 2  \right] \ ,
\end{align}
where ${\boldsymbol I}$ is an $M$-dimensional identity matrix. The
adjustment of the Lagrange multiplier $\lambda$ can be
performed in the same way as in the minimum
variance case.

\subsection{Adaptive Algorithms}

To take advantage of the proposed joint optimization approach an
\textit{on-line} modified conjugate gradient method, with one
iteration per snapshot is used to solve the resulting problem. Its
derivation is based on \cite{Chang00} and it can be interpreted as
an extension of the idea in \cite{Wang10}.

\subsubsection{Robust-CMV-MCG}

In the proposed algorithm an exponentially decayed data window is
used to estimate the matrix ${\boldsymbol R}_{xx}$ as described by
\begin{align}
\hat{{\boldsymbol R}}_{xx}(i)   =  \mu \hat{{\boldsymbol R}}_{xx}(i-1) + {\boldsymbol x}(i){\boldsymbol x}^H(i)  \ ,
\end{align}
where $\mu$ is the forgetting factor. According to \cite{Trees02}
\begin{align}
\boldsymbol{R}_{xx}       \simeq       \left[ 1-\mu \right]  \hat{\boldsymbol{R}}_{xx}(i)
\label{eq:eddw_bias}
\end{align}
can be assumed for large $i$. Replacing $\boldsymbol{R}_{xx}$ in (\ref{eq:w_rcmv}), introducing $\hat{\lambda}(i)=\frac{\lambda(i)}{1-\mu}$, leads to
${\boldsymbol w}(i) =   [ \hat{\boldsymbol {R}}_{xx}(i)  +  \tilde{\epsilon} \hat{\lambda}(i)  {\boldsymbol I}  ]^{-1} \hat{ \lambda}(i) {\boldsymbol a}/2$. Let us introduce the CG weight vector $\boldsymbol{v}(i)$ as follows $\boldsymbol{w}(i)=\boldsymbol{v}(i) \frac{\hat{\lambda}(i)}{2}$.
The conjugate gradient algorithm solves the problem by iteratively updating the CG weight vector
\begin{align}
\boldsymbol{v}(i)=\boldsymbol{v}(i-1)+ \alpha(i) \boldsymbol{p}(i),
\end{align}
where $\boldsymbol{p}(i)$ is the direction vector and $\alpha(i)$ is
the adaptive step size. One way  to realize the conjugate gradient
method performing one iteration per snapshot is the application of
the degenerated scheme \cite{Chang00}. Under this
condition the adaptive step size $\alpha(i)$ has to fulfill the
convergence bound given by
\begin{align}
0 \leq    \mathrm{E}\left\{   \boldsymbol{p}^H(i) \boldsymbol{g}(i)  \right\}   \leq     0.5~    \mathrm{E}\left\{   \boldsymbol{p}^H(i) \boldsymbol{g}(i-1) \right\}   ,
\label{eq:convergence_bound}
\end{align}
where $\mathrm{E}\left\{   \operatorname{Im}  \left\{
\boldsymbol{p}^H(i) \boldsymbol{g}(i-1) \right\} \right\} \approx 0$
and $\mathrm{E}\left\{   \operatorname{Im}  \left\{
\boldsymbol{p}^H(i) \boldsymbol{g}(i) \right\} \right\} \approx 0$
can be neglected. The negative gradient vector and
its recursive expression are considered in a similar fashion to
\cite{Chang00},\cite{Wang10} as described by
\begin{align}
\boldsymbol{g}(i) = & \boldsymbol{a}-[ \hat{\boldsymbol {R}}_{xx}(i)  +  \tilde{\epsilon} \hat{\lambda}(i)  {\boldsymbol I}  ] \boldsymbol{v}(i) \notag \\
 =& \boldsymbol{a}[1-\mu]+\mu \boldsymbol{g}(i-1)  \notag \\
& -[\boldsymbol{x} \boldsymbol{x}^H +\tilde{\epsilon}(\hat{\lambda}(i)-\mu \hat{\lambda}(i-1) )\boldsymbol{I}] \boldsymbol{v}(i-1)  \notag \\
& - \alpha(i)[ \hat{\boldsymbol {R}}_{xx}(i)  +  \tilde{\epsilon} \hat{\lambda}(i)  {\boldsymbol I}  ] \boldsymbol{p}(i)
\end{align}
Pre-multiplying with $\boldsymbol{p}^H(i)$, taking expectations on
both sides and considering $\boldsymbol{p}(i)$ uncorrelated with
$\boldsymbol{a}$, $\boldsymbol{x}(i)$ and $\boldsymbol{v}(i-1)$
leads to
\begin{align}
\mathrm{E}\left\{ \boldsymbol{p}^H(i)  \boldsymbol{g}(i)   \right\} \approx &  \mu \mathrm{E}\left\{ \boldsymbol{p}^H(i) \boldsymbol{g}(i-1)      \right\}  \notag  \\
& - \mathrm{E}\left\{ \alpha(i)\right\} \mathrm{E}\left\{   \boldsymbol{p}^H(i)[ \hat{\boldsymbol {R}}_{xx}(i)  +  \tilde{\epsilon} \hat{\lambda}(i)  {\boldsymbol I}  ] \boldsymbol{p}(i)    \right\}.
\label{eq:pHg}
\end{align}
Here it is assumed that the algorithm has converged, which implies
$\boldsymbol{a}[1-\mu]-[ \mathrm{E}\left\{\boldsymbol{x}
\boldsymbol{x}^H\right\}  +  \tilde{\epsilon}
\hat{\lambda}(i)[1-\mu]  \boldsymbol{I}
]\boldsymbol{v}(i-1)=\boldsymbol{0}$, where equation
(\ref{eq:eddw_bias}) is taken into account and $\hat{\lambda}(i)
\approx \hat{\lambda}(i-1)$. Introducing ${\boldsymbol p}_R =
[\hat{{\boldsymbol R}}_{xx}(i)+\hat{\lambda(i)}\tilde{\epsilon}
{\boldsymbol I}  ] {\boldsymbol p}(i)$, rearranging
(\ref{eq:pHg}) and inserting it into (\ref{eq:convergence_bound})
determines the step size within its boundaries as follows
\begin{align}
\alpha(i)=  \left[{\boldsymbol p}^H(i)  {\boldsymbol p}_R  \right]^{-1} \left( \mu-\eta \right){\boldsymbol p}^H(i) {\boldsymbol g}(i-1),
\end{align}
where  $0 \leq  \eta \leq 0.5$. The direction vector
is a linear combination of the previous direction vector and the
negative gradient given by
\begin{align}
\boldsymbol{p}(i+1)=\boldsymbol{p}(i)+\beta(i) \boldsymbol{g}(i),
\end{align}
where $\beta(i)$ is computed to avoid the reset procedure by
employing the Polak-Ribiere approach \cite{Luenberger}.
\begin{align}
\beta=[\boldsymbol{g}^H(i-1) \boldsymbol{g}(i-1) ]^{-1} [ \boldsymbol{g}(i)-\boldsymbol{g}(i-1)]^H \boldsymbol{g}(i)
\end{align}

The proposed algorithm, which is termed Robust-CMV-MCG, is described in Table \ref{tab:rcmv-mcg}.

Note that for the alternating algorithm to adjust
the Lagrange multiplier, we divide the update-term by $2$, if the
Lagrange multiplier is outside a predefined range, as it is
described in Table I. The application of the proposed algorithm
corresponds to a computational effort which is quadratic with the
number of sensor elements $M$.

\subsubsection{Robust-CCM-MCG}

The adaptive algorithm in {the} case of the constrained constant
modulus criterion is developed analogously to the minimum variance
case. The estimates of ${\boldsymbol R}_{a}$ and ${\boldsymbol d}$
are based on an exponentially decayed data window
 {and are given by}
\begin{eqnarray}
\hat{{\boldsymbol R}}_{\textrm{a}}(i) &  = & \mu \hat{{\boldsymbol R}}_{\textrm{a}}(i-1) +  \left|y(i)\right|^2 {\boldsymbol x}(i){\boldsymbol x}^H(i)  \\
\hat{{\boldsymbol d}}(i)   & = & \mu \hat{{\boldsymbol d}}(i-1) + {\boldsymbol x}(i) y^*(i)
\end{eqnarray}
Following the steps  {of the derivation of the MCG algorithm and
taking into account that}
\begin{align}
{\boldsymbol R}_{\textrm{a}} &  \simeq   [1-\mu]  \hat{{\boldsymbol R}}_{\textrm{a}}(i) \\
{\boldsymbol d} & \simeq    [1-\mu] \hat{{\boldsymbol d}}(i)
\end{align}
leads to the adaptive algorithm. Note that, in contrast to the CMV
case, here the beamforming weight vector is the same as the
conjugate gradient weight vector, which means ${\boldsymbol w}=[
\hat{\boldsymbol {R}}_{\textrm{a}}+   \tilde{\epsilon} \hat{\lambda}
{\boldsymbol I}  ]^{-1}  [ \gamma \hat{{\boldsymbol d}}
+\hat{\lambda} {\boldsymbol a} / 2  ] $. The negative gradient
vector and its recursive expression are defined as
\begin{align}
\boldsymbol{g}(i) = & [\gamma \hat{{\boldsymbol d}} +\hat{\lambda} {\boldsymbol a} / 2    ]-[ \hat{\boldsymbol {R}}_{\textrm{a}}+   \tilde{\epsilon} \hat{\lambda}  {\boldsymbol I}   ]  {\boldsymbol w}(i) \notag \\
=&   \mu {\boldsymbol g}(i-1)-\alpha(i){\boldsymbol p}_R
  - \left( \left|y(i)\right|^2 {\boldsymbol x}(i){\boldsymbol x}^H(i) \right) {\boldsymbol w}(i-1) \notag \\
  & +\gamma {\boldsymbol x}(i) y^*(i) + \nu \left[{\boldsymbol a} / \left(2 \tilde{\epsilon} \right) -  {\boldsymbol w}(i-1)  \right],
\end{align}
where $\nu  =  \left[\hat{\lambda}(i)-\mu \hat{\lambda}(i-1) \right] \tilde{\epsilon}$.
The proposed algorithm, which is termed Robust-CCM-MCG, is described in Table \ref{tab:rccm-mcg}.

\section{Simulations}

In this section, we present a number of simulation examples that
illustrate the performance of the proposed robust beamforming
algorithms and compare them with existing robust techniques that are
representative of the prior work in this area. A uniform linear
sensor array is used with $M=10$ sensors. Specifically, we consider
comparisons of the proposed algorithms with the loaded-SMI
\cite{Cox87}, the optimal SINR \cite{Li_book} (page 54) and the
WC-CMV in \cite{Vorobyov03}. We examine scenarios in which the SINR
is measured against the parameter $\epsilon$ that arises from the
worst-case optimization, the number of snapshots and the SNR.
 {We also consider a specific situation in which the
array steering vector is corrupted by local coherent scattering, and
scenarios in which there are changes in the environment and the
tracking performance of the beamformers is evaluated.} These
experiments are important to assess the performance of the proposed
algorithms and to illustrate how they perform against existing
methods.

\subsection{Proposed WC-CCM Algorithm}

 {In this part of the simulations, the WC-CCM design
algorithm of Table \ref{tab:wcccm} that uses a SOC program is
compared to the loaded-SMI \cite{Cox87}, the optimal SINR
\cite{Li_book} and the worst-case optimization-based constrained
minimum variance algorithm \cite{Vorobyov03}.} In the next
simulations, it is considered that $\left|s_{1}\right|=1$,
$\delta=1$, $\gamma=1$, $\epsilon=2.1$ and $\mu=0.995$ unless
otherwise specified. In addition to user $1$, the desired signal,
there are $4$ interferers, the powers $(P)$ relative to user 1 and
directions of arrival (DoA) in degrees of which are detailed in
Table \ref{tab:interference_scenario}.

The array steering vector is corrupted by local coherent scattering
\begin{align}
{\boldsymbol a}_1={\boldsymbol a}+\sum^{4}_{k=1} \mathrm e^{j
\Phi_{k}} {\boldsymbol a}_{\textrm{sc}}\left(\theta_{k}\right),
\end{align}
where $\Phi_{k}$ is uniformly distributed between zero and $2 \pi$
and $\theta_{k}$ is uniformly distributed with a standard deviation
of 2 degrees with the assumed direction as the mean. The mismatch
changes for every realization and is fixed over the snapshots of
each simulation trial.

%\begin{figure}[!htb]
%\begin{center}
%\def\epsfsize#1#2{1\columnwidth}
%\epsfbox{figure1.eps} \caption{SINR versus $\epsilon$, SNR~=~15dB,
%$M=10$, $i=200$.} \label{fig:fig1}
%\end{center}
%\end{figure}

%\begin{figure}[!htb]
%\begin{center}
%\def\epsfsize#1#2{1\columnwidth}
%\epsfbox{figure2.eps} \caption{SINR versus $\epsilon$, perfect ASV,
%$M=10$.} \label{fig:fig2}
%\end{center}
%\end{figure}

\begin{figure}[!htb]
\begin{center}
\def\epsfsize#1#2{0.65\columnwidth}
\epsfbox{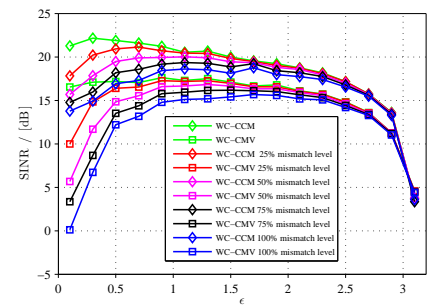} \caption{SINR versus $\epsilon$, SNR~=~15dB,
$M=10$, $i=200$.} \label{fig:fig1}
\end{center}
\end{figure}

\begin{figure}[!htb]
\begin{center}
\def\epsfsize#1#2{0.65\columnwidth}
\epsfbox{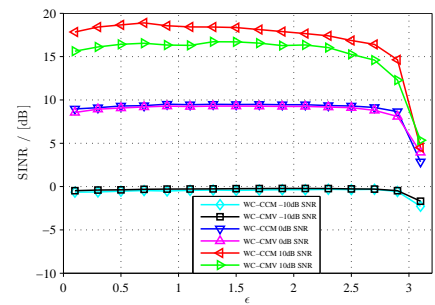} \caption{SINR versus $\epsilon$, perfect ASV,
$M=10$.} \label{fig:fig2}
\end{center}
\end{figure}

Fig.~\ref{fig:fig1} shows the SINR as a function of the design
parameter $\epsilon$ for different levels of mismatch, where its
level corresponds to the standard deviation of the local scattering.
In Fig.~\ref{fig:fig2} no mismatch is considered but different noise
levels. Both simulations show performance degradations when
$\epsilon$ is chosen close to $\sqrt{M}$, especially for high SNR
values. The simulations corroborate the analysis and show that the
optimal value for $\epsilon$ depends on the SNR.

%\begin{table} \caption{
%Interference scenario } \label{tab:interference_scenario}
%\vspace{-0.5ex}
%\centering{ $P$(dB) relative to user~1 / DoA}
%\begin{tabular}{|c|c|c|c|c|c|}
%\hline
%Snapshot & user 1  & user 2& user 3& user 4& user 5\\
% &  (desired user)& &  &  &  \\
%\hline
%$1-1000$ & $0/93^{\circ}$ & $13/120^{\circ}$ & $1/140^{\circ}$ & $22/67^{\circ}$ & $10/157^{\circ}$ \\
%$1001-2000$ & $0/93^{\circ}$ & $30/120^{\circ}$ & $25/170^{\circ}$ & $4/104^{\circ}$ & $9/68^{\circ}$ \\
%\hline
%\end{tabular}
%\end{table}

%\begin{figure}[!htb]
%\begin{center}
%\def\epsfsize#1#2{1\columnwidth}
%\epsfbox{figure3.eps} \caption{SINR versus snapshots, SNR~=~0~dB,
%local coherent scattering.} \label{fig:fig3}
%\end{center}
%\end{figure}

%\begin{figure}[!htb]
%\begin{center}
%\def\epsfsize#1#2{1\columnwidth}
%\epsfbox{figure4.eps} \caption{SINR versus SNR, local coherent
%scattering, $i=500$, $M=10$.} \label{fig:fig4}
%\end{center}
%\end{figure}

\begin{figure}[!htb]
\begin{center}
\def\epsfsize#1#2{0.65\columnwidth}
\epsfbox{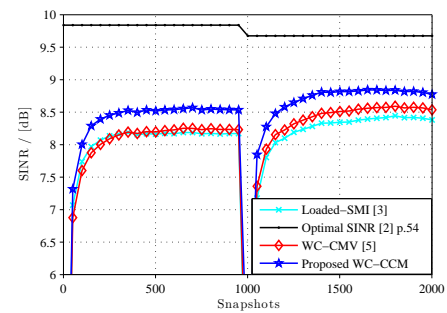} \caption{SINR versus snapshots, SNR~=~0~dB,
local coherent scattering.} \label{fig:fig3}
\end{center}
\end{figure}

\begin{figure}[!htb]
\begin{center}
\def\epsfsize#1#2{0.65\columnwidth}
\epsfbox{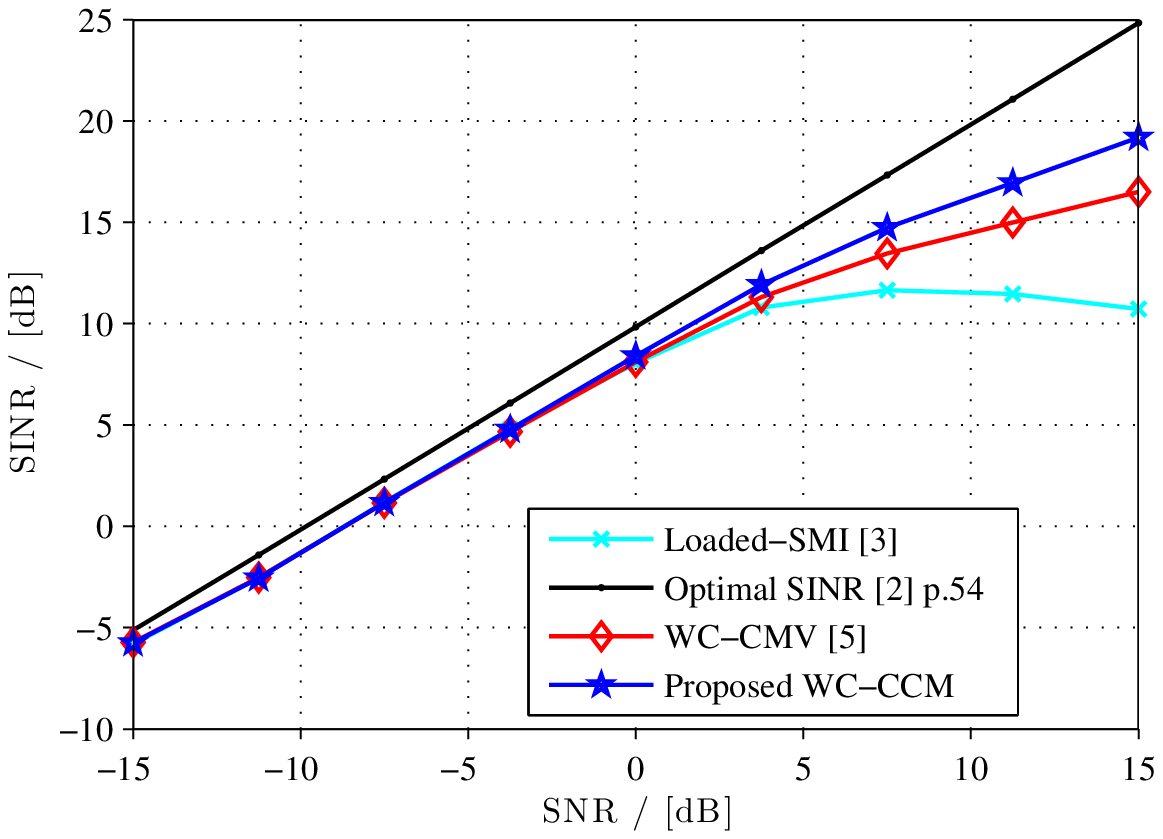} \caption{SINR versus SNR, local coherent
scattering, $i=500$, $M=10$.} \label{fig:fig4}
\end{center}
\end{figure}

Fig.~\ref{fig:fig3} presents the SINR performance over the snapshots
in the presence of local coherent scattering.  {At time index
$i=1000$ the interference scenario changes according to
Table~\ref{tab:interference_scenario} and interferers assume
different power levels and DoAs. With this change, the beamformers
must adapt to the new environment and their tracking performance is
assessed by a plot showing the SINR performance against the
snapshots.  {The proposed WC-CCM algorithm shows in
Fig.~\ref{fig:fig3} a significantly better SINR performance than the
WC-CMV \cite{Vorobyov03} and the loaded-SMI algorithm.} In terms of
tracking performance, the proposed WC-CCM algorithm of Table
\ref{tab:wcccm} is able to effectively adjust to the new
environment.} Fig.~\ref{fig:fig4} shows the SINR performance against
the SNR for $i=500$ snapshots. The curves show that the proposed
WC-CCM algorithm is more robust against mismatch problems than the
existing WC-CMV and loaded-SMI agorithms.

\subsection{Low-Complexity Robust Adaptive Beamforming}

 {In this subsection, we assess the SINR performance
of the proposed low-complexity robust beamforming algorithms in
Tables \ref{tab:rcmv-mcg} and \ref{tab:rccm-mcg} that are devised
for an online operation. In the simulations, the same parameters of
the previous subsection are used and, in addition, the step sizes
are $\mu_{\lambda}(\text{CMV})=800$ and
$\mu_{\lambda}(\text{CCM})=100$.} The limitation on the update is
set to $\delta_{\lambda max }=200$. For the robust constraints, we
employ $\epsilon=\tilde{\epsilon}=2.1$ and the parameters
$\left|s_{1}\right|=1$, $\delta=1$, $\gamma=1$,. According to the
different constraint functions, the equality is a special case for
$M=10$ and cannot be generalized. In addition to the desired user
(user 1), there are 4 interferers whose relative powers ($P$) with
respect to the desired user and directions of arrival (DoA) in
degrees are detailed in Table~\ref{tab:interference_scenario_b}.
 {At time index $i=1000$ the adaptive beamforming
algorithms are confronted with a change of scenario given in
Table~\ref{tab:interference_scenario} and the interferers assume
different power levels and DoAs. In this situation, the adaptive
beamforming algorithms must adapt to the new conditions and their
tracking performance is evaluated.}

\begin{figure}[!htb]
\begin{center}
\def\epsfsize#1#2{0.65\columnwidth}
\epsfbox{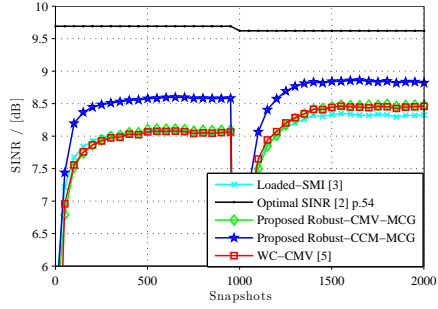} \caption{SINR versus snapshots, local coherent
scattering, SNR~=~0dB and $\left|s_{1}\right|=1$ and $\delta=1$.}
\label{fig:fig5}
\end{center}
\end{figure}

\begin{figure}[!htb]
\begin{center}
\def\epsfsize#1#2{0.65\columnwidth}
\epsfbox{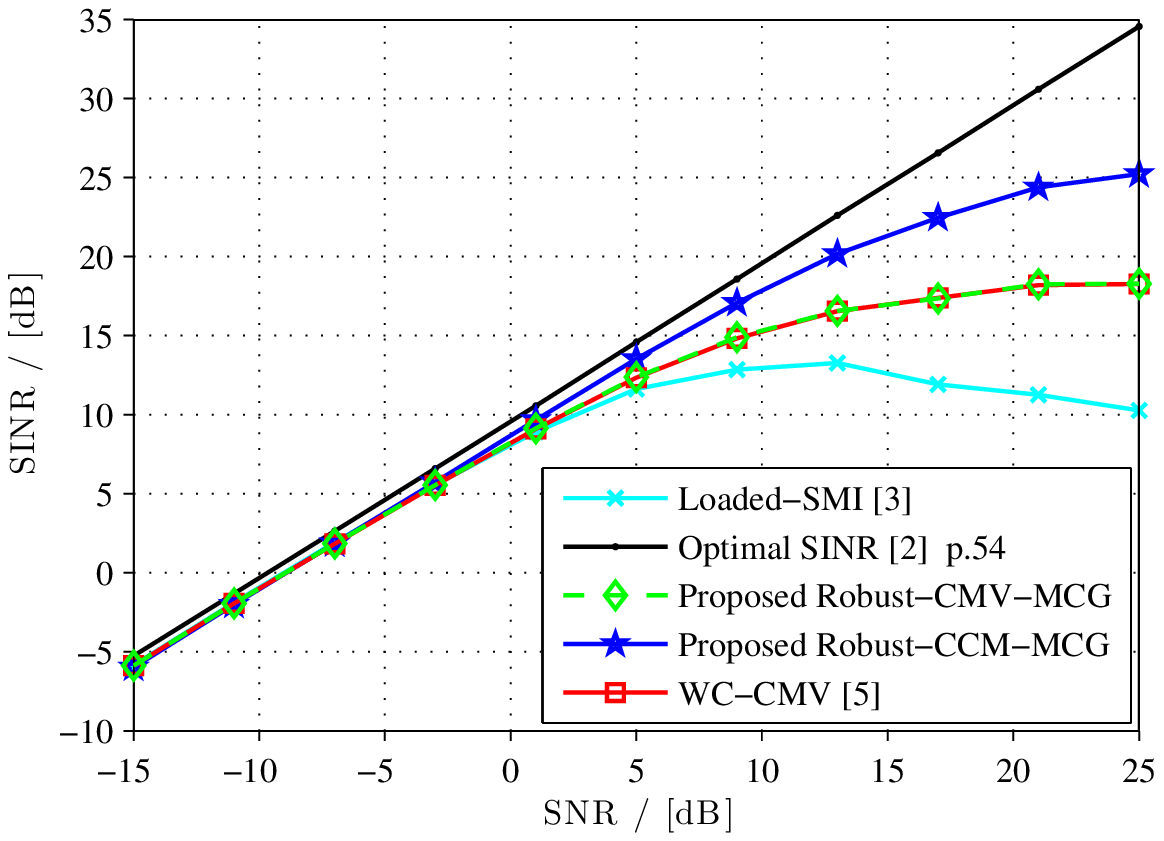} \caption{SINR versus SNR, local coherent
scattering, $i=1500$, $M=10$.} \label{fig:fig6}
\end{center}
\end{figure}

\begin{figure}[!htb]
\begin{center}
\def\epsfsize#1#2{0.65\columnwidth}
\epsfbox{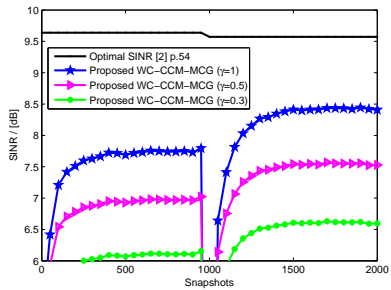} \caption{SINR versus snapshots with different
$\gamma$, local coherent scattering, SNR~=~0dB, .} \label{fig:fig7}
\end{center}
\end{figure}

Fig.~\ref{fig:fig5} shows the SINR performance as a function of the
number of snapshots in the presence of local coherent scattering.
 {The results of Fig.~\ref{fig:fig5} show that the
proposed Robust CCM-MCG algorithm has a superior SINR performance to
the existing WC-CMV \cite{Vorobyov03} algorithm, the proposed Robust
CMV-MCG algorithm and the loaded-SMI algorithm. The Robust CMV-MCG
algorithm has a comparable performance to the WC-CMV
\cite{Vorobyov03} algorithm but the latter has a significantly
higher computational cost.} The SINR performance versus the SNR is
presented in Fig.~\ref{fig:fig6}. While the proposed Robust CMV-MCG
algorithm shows an equivalent performance to the WC-CMV
\cite{Vorobyov03}, the proposed Robust CCM-MCG algorithm exploits
the constant modulus property and performs better than existing
approaches.  Fig. ~\ref{fig:fig7} shows the SINR performance against
the number of snapshots for the same scenario as in Fig.
~\ref{fig:fig5} with different values of $\gamma$ whilst keeping
delta fixed. The results show that for certain values the convexity
constraint is satisfied and the algorithm converges to a higher SINR
value, whereas for smaller values of gamma the algorithm converges
to lower values of SINR, suggesting that a local minimum of the
constant modulus cost function might have been reached. {Therefore,
the values of $\gamma$ should be set appropriately in order to
ensure an optimized performance. This adjustment could be performed
with either some prior knowledge about the energy of the signal or
with the help of a procedure that computes the energy of the signal
online. }

%\begin{table}
%\caption{Interference scenario} \label{tab:interference_scenario_b}
%\centering{ $P$(dB) relative to user 1 / DoA}
%\begin{tabular}{|c|c|c|c|c|c|}
%\hline
%Snapshot & user 1  & user 2 & user 3& user 4& user 5\\
% &  (desired user)& &  &  &  \\ \hline
%$1-1000$ & $0/93^{\circ}$ & $10/120^{\circ}$ & $5/140^{\circ}$ & $10/150^{\circ}$ & $7/105^{\circ}$ \\
%$1001-2000$ & $0/93^{\circ}$ & $30/120^{\circ}$ & $34/170^{\circ}$ & $6/104^{\circ}$ & $9/68^{\circ}$ \\
%\hline
%\end{tabular}
%\end{table}

%\begin{figure}[!htb]
%\begin{center}
%\def\epsfsize#1#2{1\columnwidth}
%\epsfbox{figure5.eps} \caption{SINR versus snapshots, local coherent
%scattering, SNR~=~0dB.} \label{fig:fig5}
%\end{center}
%\end{figure}

%\begin{figure}[!htb]
%\begin{center}
%\def\epsfsize#1#2{1\columnwidth}
%\epsfbox{figure6.eps} \caption{SINR versus SNR, local coherent
%scattering, $i=1500$, $M=10$.} \label{fig:fig6}
%\end{center}
%\end{figure}

\section{Conclusion}

We have proposed a robust beamforming algorithm based on the worst
case constraint and the constrained constant modulus
(CCM) design criterion which is called worst-case
constant modulus criterion (WC-CCM). The proposed approach exploits
the constant modulus property of the desired signal. The problem can
be solved iteratively, where each iteration is effectively solved by
a SOC program. Compared to the conventional worst-case optimization
based approach using the minimum variance design, the proposed
algorithm shows better results especially in the high SNR regime.

In addition to the WC-CCM algorithm, {we have also developed two
low-complexity robust adaptive beamforming algorithms, namely, the
Robust-CMV-MCG and the Robust-CCM-MCG}. The proposed algorithms use
a constraint similar to the worst-case optimization based approach.
It has been shown that the joint optimization approach allows the
exploitation of highly efficient \textit{on-line} algorithms like
the modified conjugate gradient method which performs just one
iteration per snapshot taking advantage of previous computations. As
a result the complexity is reduced by more than an order of
magnitude compared to the worst-case optimization based beamformer
which is solved with a second-order cone program. While the proposed
Robust-CMV-MCG performs  {equivalently}, the proposed Robust-CCM-MCG
algorithm based on the CCM design criterion, shows a better
performance which takes advantage of the constant modulus property
of the signal amplitude of the desired user.

\vspace{2ex}

\bibliographystyle{IEEEtran}
%\biboptions{semicolon}
\bibliography{refs_ltnl_Jul31}
%\begin{biography}[{\includegraphics[width=1in,height=1.25in,clip,keepaspectratio]{mshell}}]{Michael Shell}

%\end{IEEEbiography}


%\newpage



%\vfill


\newpage
\section*{Figures}

\newpage

\section*{Tables}

\begin{table}[!htb]
\caption{Proposed WC-CCM Algorithm}
\vspace{1em}
\label{tab:wcccm}
\centering{
\begin{tabular}{|l|}
\hline
\\[-2.2ex]
initialization: $\hat{{\boldsymbol R}}_{a}(0)=\sigma_{n}^{2} \boldsymbol{I}; \hat{\boldsymbol d}(0)=\boldsymbol{0}; \boldsymbol{w}(0)=\frac{\boldsymbol{a}}{M}$ \\ \\[-2.2ex]
Update for each time instant i = 1,...,N \\ \\[-2.2ex]
~~~$y(i)=\boldsymbol{w}^H(i-1) \boldsymbol{x}(i) $ \\ \\[-2.2ex]
~~~$\hat{{\boldsymbol R}}_{a}(i)=\mu \hat{{\boldsymbol R}}_{a}(i-1)+ \left|y(i)\right|^2  {\boldsymbol x}(i){\boldsymbol x}^H(i) $ \\ \\[-2.2ex]
~~~${\boldsymbol R}_{\textrm{ac}}(i)=  \mathrm{chol}\left( \hat{{\boldsymbol R}}_{a}(i)  \right)    $\\ \\[-2.2ex]
~~~$\hat{\boldsymbol d}(i)=\mu \hat{\boldsymbol d}(i-1) + {\boldsymbol x}(i) y^{*}(i)$  \\ \\[-2.2ex]
~~~${\boldsymbol R}_{\textrm{acr}}(i)=  \begin{bmatrix} \operatorname{Re} \left\{ {\boldsymbol R}_{\textrm{ac}}(i) \right\} &     -\operatorname{Im} \left\{ {\boldsymbol R}_{\textrm{ac}}(i) \right\}  \\ \operatorname{Im} \left\{ {\boldsymbol R}_{\textrm{ac}}(i) \right\}         & \operatorname{Re} \left\{ {\boldsymbol R}_{\textrm{ac}}(i) \right\}  \end{bmatrix}  $\\ \\[-2.2ex]
~~~${\boldsymbol d}_{\textrm{r}}(i)= \left[ \operatorname{Re}\left\{ \hat{\boldsymbol d}(i)  \right\}^T ,   \operatorname{Im} \left\{ \hat{\boldsymbol d}(i)  \right\}^T  \right]^T $\\ \\[-2.2ex]
~~~${\boldsymbol p}  =  [ 1, {\boldsymbol 0}^T ]^T  $ \\
~~~${\boldsymbol f}  =  [ 1/2 , 1/2 ,  {\boldsymbol 0}^T,-\delta,{\boldsymbol 0}^T,0 ]^T   $\\ \\[-2.2ex]
~~~${\boldsymbol F}^T  =  \begin{bmatrix}   \frac{1}{2} & \gamma {\boldsymbol d}_{\textrm{r}}^T(i) \\ -\frac{1}{2} & - \gamma {\boldsymbol d}_{\textrm{r}}^T(i) \\ \\[-2.2ex] {\boldsymbol 0} & {\boldsymbol R}_{\textrm{acr}}(i) \\ 0 & \breve{{\boldsymbol a}} \\ \\[-2.2ex] {\boldsymbol 0} & \epsilon {\boldsymbol I} \\ 0 & \bar{{\boldsymbol a}}  \end{bmatrix}$ \\ \\ \\[-2.2ex]
~~~$ \min\limits_{{\boldsymbol u}}~~{\boldsymbol p}^T {\boldsymbol u} ~~
  {\rm s.~t.~} $  \\ \\[-2.2ex]
~~~~~~~~~~~~~~$  {\boldsymbol f}+ {\boldsymbol F}^T {\boldsymbol u}~~ \in  \mathrm{SOC}_{1}^{2M+2} \times  \mathrm{SOC}_{2}^{2M+1}   \times \{0 \} $ \\ \\ \\[-2.2ex]
~~~${\boldsymbol w}(i)= \left[ {\boldsymbol u}_2,...,{\boldsymbol u}_{M+1}\right]^T  + j \left[ {\boldsymbol u}_{M+2},...,{\boldsymbol u}_{2M+1}\right]^T$\\ \\ \hline
\end{tabular}
}
\end{table}

\begin{table}[!htb]
\caption{Proposed RCMV-MCG Algorithm}
\vspace{1em}
\label{tab:rcmv-mcg}
\centering{
\begin{tabular}{|l|}
\hline
\\[-2.2ex]
${\boldsymbol v}(0)={\boldsymbol 0}; \  {\boldsymbol p}(1)={\boldsymbol g}(0)={\boldsymbol a}    ; \ \hat{{\boldsymbol R}}(0)= \delta {\boldsymbol I}; \ \hat{\lambda}(0)=\hat{\lambda}(1)=\hat{\lambda}_{0} $   \\ \\[-2.2ex]
\textbf{For each time instant} $i=1,...,N$ \\ \\[-2.2ex]
~~~$\hat{{\boldsymbol R}}_{xx}(i)=\mu \hat{{\boldsymbol R}}_{xx}(i-1)+ {\boldsymbol x}(i){\boldsymbol x}^H(i) $ \\ \\[-2.2ex]
~~~${\boldsymbol p}_R = [\hat{{\boldsymbol R}}_{xx}(i)+\hat{\lambda}(i)\tilde{\epsilon} {\boldsymbol I}  ] {\boldsymbol p}(i)$; \ \ $\nu  =  \left[\hat{\lambda}(i)-\mu \hat{\lambda}(i-1) \right] \tilde{\epsilon}$ \\ \\[-2.2ex]
~~~$\alpha(i)=  \left[{\boldsymbol p}^H(i)   {\boldsymbol p}_R  \right]^{-1} \left( \mu-\eta \right){\boldsymbol p}^H(i) {\boldsymbol g}(i-1)$; \ \
$\left( 0 \leq  \eta \leq 0.5 \right) $ \\ \\[-2.2ex]
~~~${\boldsymbol v}(i)={\boldsymbol v}(i-1) + \alpha(i) {\boldsymbol p}(i)$\\  \\[-2.2ex]
~~~${\boldsymbol g}(i)=[1-\mu]{\boldsymbol a}+\mu {\boldsymbol g}(i-1)-\alpha(i){\boldsymbol p}_R $\\   \\[-2.2ex]
~~~~~~~~~~~~$- \left(  {\boldsymbol x}(i){\boldsymbol x}^H(i) +\nu {\boldsymbol I} \right) {\boldsymbol v}(i-1)  $\\  \\[-2.2ex]
~~~$\beta(i)=\left[  {\boldsymbol g}^H(i-1) {\boldsymbol g}(i-1)  \right]^{-1}    \left[  {\boldsymbol g}(i) - {\boldsymbol g}(i-1)   \right]^H {\boldsymbol g}(i)$\\   \\[-2.2ex]
~~~${\boldsymbol p}(i+1)={\boldsymbol g}(i)+\beta(i){\boldsymbol p}(i) $\\  \\[-2.2ex]
~~~${\boldsymbol w}(i)=\lambda(i) {\boldsymbol v}(i)/2 $\\ \\
~~~$\delta_{\lambda}=\mu_{\lambda}  [ \tilde{\epsilon} \left\|{\boldsymbol w}(i)\right\|^{2}_{2} -\operatorname{Re}\left\{{\boldsymbol w}^H(i)    \boldsymbol{a}  \right\}   + \delta ] $\\  \\[-2.2ex]
~~~while $\delta_{\lambda}\leq-\lambda(i)$ or $\delta_{\lambda}\geq \delta_{\lambda \textrm{max}}$\\  \\[-2.2ex]
~~~~~~$\delta_{\lambda} \Rightarrow  \delta_{\lambda}/2   $\\  \\[-2.2ex]
~~~end \\ \\[-2.2ex]
~~~$ \hat{\lambda}(i+1)=\hat{\lambda}(i)+ \delta_{\lambda} $\\ \\
\hline
\end{tabular}}
\end{table}

\begin{table}[!htb]
\caption{Proposed RCCM-MCG Algorithm}
\vspace{1em}
\label{tab:rccm-mcg}
\centering{
\begin{tabular}{|l|}
\hline
\\[-2.2ex]
$ {\boldsymbol p}(1)={\boldsymbol g}(0)={\boldsymbol a}  ; \ \hat{{\boldsymbol R}}_{\textrm{a}}(0)= \delta {\boldsymbol I}; \ \hat{\boldsymbol d}(0)={\boldsymbol 0}$; \\ $\hat{\lambda}(0)=\hat{\lambda}(1)=\hat{\lambda}_{0}; \   {\boldsymbol w}={\boldsymbol a}/M  $  \\ \\[-2.2ex]
\textbf{For each time instant} $i=1,...,N$ \\ \\[-2.2ex]
~~~$\hat{{\boldsymbol R}}_{\textrm{a}}(i)=\mu \hat{{\boldsymbol R}}_{\textrm{a}}(i-1)+ \left|y(i)\right|^2 {\boldsymbol x}(i){\boldsymbol x}^H(i) $\\ \\[-2.2ex]
~~~${\boldsymbol p}_R = [\hat{{\boldsymbol R}}_{\textrm{a}}(i)+\hat{\lambda}(i) \tilde{\epsilon} \boldsymbol{I} ]   {\boldsymbol p}(i)$; \ \ $\nu  =  \left[\hat{\lambda}(i)-\mu \hat{\lambda}(i-1) \right] \tilde{\epsilon}$ \\ \\[-2.2ex]
~~~$\alpha(i)=  \left[{\boldsymbol p}^H(i)  {\boldsymbol p}_R   \right]^{-1} \left( \mu-\eta \right){\boldsymbol p}^H(i) {\boldsymbol g}(i-1); \ \
\left( 0 \leq  \eta \leq 0.5 \right)$  \\ \\[-2.2ex]
~~~${\boldsymbol w}(i)={\boldsymbol w}(i-1) + \alpha(i) {\boldsymbol p}(i)$\\ \\[-2.2ex]
~~~${\boldsymbol g}(i) =  \mu {\boldsymbol g}(i-1)-\alpha(i){\boldsymbol p}_R
  - \left( \left|y(i)\right|^2 {\boldsymbol x}(i){\boldsymbol x}^H(i) \right) {\boldsymbol w}(i-1) $\\ \\[-2.2ex]
~~~~~~~~~~~~$ +\gamma {\boldsymbol x}(i) y^*(i) + \nu \left[{\boldsymbol a} / \left(2 \tilde{\epsilon} \right) -  {\boldsymbol w}(i-1)  \right]$  \\ \\[-2.2ex]
~~~$\beta(i)=\left[  {\boldsymbol g}^H(i-1) {\boldsymbol g}(i-1)  \right]^{-1}    \left[  {\boldsymbol g}(i) - {\boldsymbol g}(i-1)   \right]^H {\boldsymbol g}(i)$\\ \\[-2.2ex]
~~~${\boldsymbol p}(i+1)={\boldsymbol g}(i)+\beta(i){\boldsymbol p}(i) $\\ \\
~~~$\delta_{\hat{\lambda}}=\mu_{\hat{\lambda}}  [ \hat{\tilde}{\epsilon} \left\|{\boldsymbol w}(i)\right\|^{2}_{2} -\operatorname{Re}\left\{{\boldsymbol w}^H(i)    \boldsymbol{a}  \right\}   + \delta ] $\\ \\[-2.2ex]
~~~while $\delta_{\lambda}\leq-\hat{\lambda}(i)$ or $\delta_{\lambda}\geq \delta_{\lambda \textrm{max}}$\\ \\[-2.2ex]
~~~~~~$\delta_{\lambda} \Rightarrow  \delta_{\lambda}/2   $\\ \\[-2.2ex]
~~~end \\ \\[-2.2ex]
~~~$ \hat{\lambda}(i+1)=\hat{\lambda}(i)+ \delta_{\lambda} $\\ \\
\hline
\end{tabular}}
\end{table}

\begin{table}[!htb]
\caption{Interference scenario } \label{tab:interference_scenario}
%\vspace{-0.5ex}
\centering{ $P$(dB) relative to user1 / DoA}
\begin{tabular}{|c|c|c|c|c|c|}
\hline
Snapshot & user 1  & user 2& user 3& user 4& user 5\\
 &  (desired user)& &  &  &  \\
\hline
1-1000& 0/93° &13/120° &1/140° &22/67° & 10/157°\\
1001-2000 & 0/93°& 30/120° &25/170°& 4/104° &9/68°\\
\hline
\end{tabular}
\end{table}

\begin{table}[!htb]
\caption{Interference scenario}
\label{tab:interference_scenario_b}
\centering{ $P$(dB) relative to user1 / DoA}
\begin{tabular}{|c|c|c|c|c|c|}
\hline
Snapshot & user 1  & user 2& user 3& user 4& user 5\\
 &  (desired user)& &  &  &  \\ \hline
1-1000& 0/93° &10/120° &5/140° &10/150° & 7/105°\\
1001-2000 & 0/93°& 30/120° &34/170°& 6/104° &9/68°\\
\hline
\end{tabular}
\end{table}
\end{document}